\documentclass{article}
\usepackage{etoolbox}
\newtoggle{arxiv}
\toggletrue{arxiv}

\iftoggle{arxiv}{
\usepackage[nonatbib,final]{neurips_arxiv}
}{
\usepackage[nonatbib]{neurips_2019}
}

\newtoggle{release}
\toggletrue{release}

\usepackage[numbers]{natbib}
\usepackage[utf8]{inputenc}
\usepackage[T1]{fontenc}
\usepackage{amsfonts}       %
\usepackage{amsmath, amssymb, amsthm}
\usepackage{booktabs}       %
\usepackage{dsfont}
\usepackage{enumerate}
\usepackage{graphicx}
\usepackage{hyperref}
\usepackage{microtype}      %
\usepackage{nicefrac}       %
\usepackage{pgfplots}
\usepackage{pgf}
\usepackage{tikzscale}
\usepackage{tikz}
\usepackage{url}
\usepackage{xcolor}
\usepackage{subcaption}
\usepackage{ifthen}
\usepackage{pifont}
\usepackage{makecell}

\usepackage{pgfplotstable}
\pgfplotsset{compat=newest}
\usepgfplotslibrary{external}
\usetikzlibrary{external,arrows,calc,positioning,shapes.geometric,math}

\nottoggle{release}{
\newcommand{\nico}[1]{{\color{blue}#1}}
\newcommand{\alex}[1]{{\color{purple}#1}}
\newcommand{\leon}[1]{{\color{green}#1}}
\newcommand{\francis}[1]{{\color{orange}#1}}
}{
\newcommand{\nico}[1]{}
\newcommand{\neil}[1]{}
\newcommand{\alex}[1]{}
\newcommand{\leon}[1]{}
\newcommand{\francis}[1]{}
}

\DeclareMathOperator{\Conv}{Conv1d}
\DeclareMathOperator{\Convtwo}{Conv2d}
\DeclareMathOperator{\Convtr}{ConvTr1d}
\DeclareMathOperator{\ReLU}{Relu}
\DeclareMathOperator{\GLU}{GLU}
\DeclareMathOperator{\SDR}{SDR}
\newcommand{\ext}{\mathrm{ext}}
\newcommand{\sob}{\mathbf{s}}
\DeclareMathOperator{\SIR}{SIR}
\DeclareMathOperator{\SAR}{SAR}

\providecommand{\norm}[1]{\left\|#1\right\|}

\providecommand{\prob}[1]{\mathbb{P}\left\{#1\right\}}
\providecommand{\reel}{\mathbb{R}}

\newcommand{\chmark}{\ding{51}}
\newcommand{\crmark}{\ding{55}}
\newcommand{\source}[1]{\texttt{#1}}

\title{Demucs: Deep Extractor for Music Sources with extra unlabeled data remixed}

\author{
  Alexandre D\'efossez \\
  Facebook AI Research \\
  INRIA / \'Ecole Normale Sup\'erieure\\
  PSL Research University\\
  Paris, France\\
  \texttt{defossez@fb.com}
  \And
  Nicolas Usunier\\
  Facebook AI Research \\
  Paris, France\\
  \texttt{usunier@fb.com}
  \AND
  L\'eon Bottou\\
  Facebook AI Research \\
  New York, USA\\
  \texttt{leonb@fb.com}
  \And
  Francis Bach\\
  INRIA / \'Ecole Normale Sup\'erieure\\
  PSL Research University\\
  Paris, France\\
  \texttt{francis.bach@ens.fr}
}

\begin{document}

\maketitle

\begin{abstract}
  We study the problem of source separation
  for music using deep learning with four known sources: drums, bass,
  vocals and other accompaniments.
  State-of-the-art approaches predict soft masks over mixture spectrograms
  while methods working on the waveform are lagging behind as measured on the
  standard MusDB~\cite{musdb} benchmark.
  Our contribution is two fold. (i) We introduce a simple convolutional
  and recurrent model that outperforms the state-of-the-art model on waveforms, that is, Wave-U-Net~\cite{wavunet},
  by 1.6 points of SDR (signal to distortion ratio).
  (ii)~We propose a new scheme to leverage unlabeled music.
  We train a first model to extract
  parts with at least one source silent in unlabeled tracks, for instance
  without bass. We remix this extract with a bass line taken from the
  supervised dataset to form a new weakly supervised training example. Combining our architecture and scheme, we show
  that waveform methods can play in the same ballpark as spectrogram ones.

\end{abstract}

\section{Introduction}

Cherry first noticed the ``cocktail party effect''~\cite{cocktail}: how the human brain is able to
separate a single conversation out of a surrounding noise
from a room full of people chatting.
Bregman later tried to understand how the brain was
able to analyse a complex auditory signal and segment it into higher level streams.
His framework for auditory scene analysis~\cite{asa} spawned its
computational counterpart, trying to reproduce or model accomplishment
of the brains with algorithmic means~\cite{casa}.

When producing music, recordings of individual instruments called \emph{stems}
are arranged together and mastered into the final song. The goal of source
separation is then to recover those individual stems from the mixed signal.
Unlike the cocktail problem, there is not a single source of interest to
differentiate from an unrelated background noise, but instead a wide variety
of tones and timbres playing in a coordinated way. As part of the SiSec Mus
evaluation campaign for music separation~\cite{sisec}, a choice was made to regroup
those individual stems into 4 broad categories: (1)~\source{drums}, (2)~\source{bass}, (3)~\source{other}, (4)~\source{vocals}.

Each source is represented by a waveform $s_i \in \reel^{C, T}$
where $C$ is the number of channels (1 for mono, 2 for stereo) and $T$
the number of samples. We define $\sob := (s_i)_i$ the concatenation of sources in a tensor of size $(4, C, T)$ and
the mixture $s := \sum_{i=1}^4 s_i$.  We aim at training a model
that minimises
\begin{equation}
    \label{eq:separation}
    \min_\theta \sum_{\sob\in\mathcal{D}} \sum_{i=1}^4 l(g^i_\theta(s), s_i)
\end{equation}
for some dataset $\mathcal{D}$, reconstruction error $l$, model architecture $g$
with 4 outputs $g^i$, and model weights $\theta \in \reel^d$.

As presented in the next section,
most methods to solve~\eqref{eq:separation} learn a mask per source $\sigma_i$ on the mixture spectrogram $S:=\mathrm{STFT}(s)$ (Short-Time Fourier Transform).
The estimated sources are then $\hat{s}_i := \mathrm{ISTFT}(\sigma_i S)$ (Inverse Short-Time Fourier Transform). The mask $\sigma_i$ can either be a binary mask valued in $\{0, 1\}$
or a soft assignment valued in $[0, 1]$. Those methods are state-of-the-art and perform very well without requiring large models.
However, they come with two limitations:
\vspace{-0.15cm}
\begin{enumerate}
    \item There is no reason for $\sigma_i S$ to be a real spectrogram (i.e., obtained from a real signal). In that case the $\mathrm{ISTFT}$
    step will perform a projection step that is not accounted for in the training loss and could result in artifacts.
    \item Such methods do not try to model the phase but reuse the input mixture. Let us imagine that a guitar plays with a singer at the same pitch
    but the singer is doing a slight vibrato, i.e., a small modulation
    of the pitch. This modulation will impact the spectrogram phase, as the derivative
    of the phase is the instant frequency. Let say both the singer and the guitar have the same intensity, then the ideal mask would be 0.5 for each.
    However, as we reuse for each source the original phase, the vibrato from the singer would also be applied to the guitar.
    While this could be consider a corner case, its existence is a motivation for the search of an alternative.
\end{enumerate}
\vspace{-0.15cm}
Learning a model from/to the waveform could allow to lift some of the aforementioned limitations.
Because a waveform is directly generated, the training loss is end-to-end, with no extra
synthesis step that could add artifacts which solves the first point above.
As for the second point, it is unknown whether any model could succeed
in separating such a pathological case.
In the fields of speech or music generation, direct waveform synthesis has replaced spectrogram based methods~\cite{wavenet,samplernn,sing}.
When doing generation without an input signal $s$, the first point is more problematic. Indeed, there is no input phase
to reuse and the inversion of a power spectrogram will introduce significant artifacts~\cite{nsynth}. Those successes were also made possible
by the development of very large scale datasets (30GB for the NSynth dataset~\cite{nsynth}). In comparison the standard MusDB dataset is
only a few GB. This explains, at least partially, the worse performance of waveform methods for source separation~\cite{sisec}.

In this paper we aim at taking waveform based methods one step closer to spectrogram methods. We contribute a simple model architecture
inspired by previous work in source separation from the waveform and
audio synthesis. We show that this model outperforms the previous
state of the art on the waveform domain.
Given the limited data available, we further refine the performance
of our model by using a novel semi-supervised
data augmentation scheme that allows to leverage 2,000 unlabeled songs.

\section{Related Work}

A first category of methods for supervised music source separation work
on power spectrograms. They predict a power spectrogram for each source
and reuse the phase from the input mixture to synthesise individual waveforms.
Traditional methods have mostly focused on blind (unsupervised) source separation.
Non-negative matrix factorization techniques~\cite{nmf_unified} model the power spectrum
as a weighted sum of a learnt spectral dictionary, whose elements can then be
grouped into individual sources.
Independent component analysis~\cite{ica} relies on independence assumptions
and multiple microphones to separate the sources.
Learning a soft/binary mask over power spectrograms has been done using
either HMM-based prediction~\cite{one_mic_hmm} or segmentation techniques~\cite{bach2005blind}.

With the development of deep learning, fully supervised methods have gained momentum.
Initial work was performed on speech source separation~\cite{classif_dnn_separation} then
for music using simple fully connected networks over few spectrogram frames~\cite{uhlich2015deep},
LSTMs~\cite{uhlich2017improving}, or multi scale convolutional /
recurrent networks~\cite{liu2018denoising,sony_densenet,sony_denselstm}. State-of-the-art performance
is obtained with those models when trained with extra labeled data. We show that our model architecture
combined with our semi-supervised scheme can provide performance almost on par, while being trained on
5 times less labeled data.

On the other hand, working directly on the waveform only became possible with
deep learning models. A Wavenet-like but regression based approach was first used
for speech denoising~\cite{wavenet_denoising} and then adapted to source
separation~\cite{wavenet_sep}. Concurrently, a convolutional network with a U-Net
structure called Wave-U-Net was used first on spectrograms~\cite{waveunet_singing} and then
adapted to the waveform domain~\cite{wavunet}. Those methods performs significantly worse
than the spectrogram ones as shown in the latest SiSec Mus source separation evaluation campaign~\cite{sisec}.
As shown in Section~\ref{sec:results}, we
outperform Wave-U-Net by a large margin with our architecture alone.

In~\cite{lior_semi_source_separation}, the problem of semi-supervised source separation is tackled for 2 sources separation
where a dataset of mixtures and unaligned isolated examples of source 1 but not source 2 is available. Using specifically crafted adversarial losses
the authors manage to learn a separation model. In~\cite{fb_unsup_source_separation}, the case of blind, i.e., completely unsupervised
source separation is covered, combining adversarial losses with a remixing trick similar in spirit to our unlabeled data remixing presented
in Section~\ref{sec:remix}. Both papers are different from our own setup, as they assume that they completely lack isolated audio for some or all sources.
Finally, when having extra isolated sources, previous work showed that it was possible to use adversarial losses
to leverage them without using them to generate new mixtures~\cite{adversarial_semi_sup}. Unfortunately, extra isolated sources
are exactly the kind of data that is hard to come by. As far as we know, no previous work tried to leverage unlabeled songs
in order to improve supervised source separation performance. Besides, most previous work relied on adversarial losses, which can prove expensive while our remixing trick allows for direct supervision of the training loss.

\section{Model Architecture}
\label{sec:model}

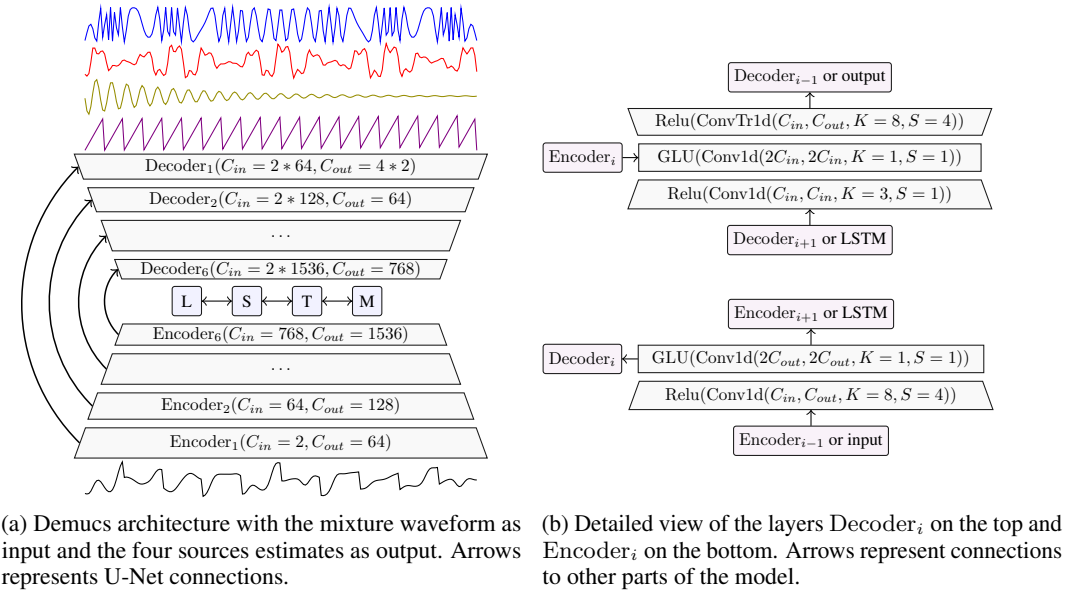
\begin{figure}
  \centering
  \def\pscale{0.65}
  \begin{subfigure}[b]{0.49\textwidth}
    \begin{tikzpicture}[
    every node/.style={scale=\pscale},
    conv/.style={shape=trapezium,
        trapezium angle=70, draw, inner xsep=0pt, inner ysep=0pt,
        draw=black!90,fill=gray!5},
    deconv/.style={shape=trapezium,
        trapezium angle=-70, draw, inner xsep=0pt, inner ysep=0pt,
        draw=black!90,fill=gray!5},
    rnn/.style={rounded corners=1pt,rectangle,draw=black!90,
        fill=blue!5,minimum width=0.6cm, minimum height=0.6cm},
    skip/.style={line width=0.2mm, ->},
]
    \def\yshift{0.3em}
    \def\base{9cm}
    \def\dec{0.55cm}

    \node (base) at (0, 0) {};
    \def\sourcea{0.25 * 2 * (0.5 * x - floor(0.5 * x))}
    \def\sourceb{0.25 * exp(-(x + 10)/10) * cos(deg(4 * x))}
    \def\sourcec{0.25 * cos(deg(0.3 * x)) * cos(deg(2 * x + 0.9 * cos(deg(4 * x)))}
    \def\sourced{0.25 * cos(deg(5 * x) * (
        1 + 0.1 * cos(deg(1 * x)))
     )}
    \begin{axis}[
        anchor=north,
        at=(base),
        scale=0.6,
        domain=-10:10,
        axis y line=none,
        axis x line=none,
        samples=200,
        color=black,
        height=2.5cm,
        width=\base + 3cm]
          \addplot[mark=none] {
              (
                \sourcea + \sourceb + \sourcec + \sourced
               )
         };
    \end{axis}
    \node (e1) [conv, minimum width=\base - \dec, anchor=south] at (0, 0)
        {$\mathrm{Encoder}_1(C_{in}=2, C_{out}=64)$};
    \node (e2) [conv, minimum width=\base - 2*\dec, anchor=south] at
        ($(e1.north) + (0, \yshift)$)
        {$\mathrm{Encoder}_2(C_{in}=64, C_{out}=128)$};
    \node (edots) [conv, minimum width=\base - 3*\dec, anchor=south] at
        ($(e2.north) + (0, \yshift)$)
        {$\ldots$};
    \node (e6) [conv, minimum width=\base - 4*\dec, anchor=south] at
        ($(edots.north) + (0, \yshift)$)
        {$\mathrm{Encoder}_6(C_{in}=768, C_{out}=1536)$};

    \node (ls0) [rnn] at ($(e6.north) + (-0.35 * \base + 3 * \dec + 0.25cm,0.3cm)$) {L};
    \foreach \k/\text in {1/S,2/T,3/M} {
        \tikzmath{
            int \prev;
            \prev=\k - 1;
        }
        \node (ls\k) [rnn,anchor=west] at ($(ls\prev.east) + (0.4cm, 0)$) {\text};
        \draw [<->] (ls\prev) -- (ls\k);
    }
    \node (d6) [deconv, minimum width=\base - 4*\dec, anchor=south] at
        ($(e6.north) + (0, 0.59cm)$) {$\mathrm{Decoder}_6(C_{in}=2 * 1536, C_{out}=768)$};
    \node (ddots) [deconv, minimum width=\base - 3*\dec, anchor=south] at
        ($(d6.north) + (0, \yshift)$) {$\ldots$};
    \node (d2) [deconv, minimum width=\base - 2*\dec, anchor=south] at
        ($(ddots.north) + (0, \yshift)$) {$\mathrm{Decoder}_2(C_{in}=2 * 128, C_{out}=64)$};
    \node (d1) [deconv, minimum width=\base - \dec, anchor=south] at
        ($(d2.north) + (0, \yshift)$) {$\mathrm{Decoder}_1(C_{in}=2 * 64, C_{out}=4 * 2)$};

    \path[skip] (e1.west) edge[bend left=45] node [right] {} (d1.west);
    \path[skip] (e2.west) edge[bend left=45] node [right] {} (d2.west);
    \path[skip] (edots.west) edge[bend left=45] node [right] {} (ddots.west);
    \path[skip] (e6.west) edge[bend left=45] node [right] {} (d6.west);
    \newcommand\myoutput[3]{
        \begin{axis}[
            anchor=south,
            scale=0.6,
            at=#1,
            domain=-20:20,
            axis y line=none,
            axis x line=none,
            samples=200,
            height=2.5cm,
            color=#2,
            width=\base + 3cm]
            \addplot[mark=none] {
                #3
            };
        \end{axis}
    }
    \node (o1) at (d1.north) {};
    \node (o2) at ($(o1.north) + (0, 4mm)$) {};
    \node (o3) at ($(o2.north) + (0, 4mm)$) {};
    \node (o4) at ($(o3.north) + (0, 4mm)$) {};
    \myoutput{(o1)}{violet}{\sourcea}
    \myoutput{(o2)}{olive}{\sourceb}
    \myoutput{(o3)}{red}{\sourcec}
    \myoutput{(o4)}{blue}{\sourced}
\end{tikzpicture}

    \caption{Demucs architecture with the mixture waveform as input
    and the four sources estimates as output. Arrows represents U-Net
    connections.}
  \end{subfigure}%
  \hfill%
  \begin{subfigure}[b]{0.49\textwidth}
    \begin{tikzpicture}[
    every node/.style={scale=\pscale},
    conv/.style={shape=trapezium,
        trapezium angle=70, draw, inner xsep=0pt, inner ysep=2pt,
        draw=black!90,fill=gray!5},
    deconv/.style={shape=trapezium,
        trapezium angle=-70, draw, inner xsep=0pt, inner ysep=2pt,
        draw=black!90,fill=gray!5},
    rewrite/.style={shape=rectangle,
        draw, inner xsep=11pt, inner ysep=3pt,
        draw=black!90,fill=gray!5},
    inout/.style={rounded corners=1pt,rectangle,draw=black!90,
        fill=violet!5,minimum width=0.6cm, minimum height=0.6cm},
    skip/.style={line width=0.2mm, ->},
]
    \def\yshift{0.3em}
    \def\base{8cm}
    \def\dec{0.55cm}

    \node (base) at (0, 0cm) {};
    \node (pad) at (0, -1cm) {};
    \node (conv) [conv, minimum width=\base - \dec, anchor=south] at
      (base.north) {$\ReLU(\Conv(C_{in}, C_{in}, K=3, S=1))$};
    \node (rewrite) [rewrite, minimum width=\base - 2 * \dec, anchor=south] at
      ($(conv.north) + (0,\yshift)$) {$\GLU(\Conv(2 C_{in}, 2 C_{in}, K=1, S=1))$};
    \node (deconv) [deconv, minimum width=\base - \dec, anchor=south] at
      ($(rewrite.north) + (0,\yshift)$) {$\ReLU(\Convtr(C_{in},C_{out}, K=8, S=4))$};

    \def\yshift{0.6em}
    \node (skip) [inout, anchor=east] at ($(rewrite.west) - (\yshift, 0)$) {$\mathrm{Encoder}_i$};
    \draw[->]  (skip) -- (rewrite.west);

    \node (prev) [inout, anchor=north] at ($(conv.south) - (0, \yshift)$) {$\mathrm{Decoder}_{i+1}$ or LSTM};
    \draw[->]  (prev.north) -- (conv.south);

    \node (next) [inout, anchor=south] at ($(deconv.north) + (0, \yshift)$) {$\mathrm{Decoder}_{i-1}$ or output};
    \draw[->]  (deconv.north) -- (next.south);
\end{tikzpicture}

    \begin{tikzpicture}[
    every node/.style={scale=\pscale},
    conv/.style={shape=trapezium,
        trapezium angle=70, draw, inner xsep=0pt, inner ysep=2pt,
        draw=black!90,fill=gray!5},
    deconv/.style={shape=trapezium,
        trapezium angle=-70, draw, inner xsep=0pt, inner ysep=2pt,
        draw=black!90,fill=gray!5},
    rewrite/.style={shape=rectangle,
        draw, inner xsep=8pt, inner ysep=3pt,
        draw=black!90,fill=gray!5},
    inout/.style={rounded corners=1pt,rectangle,draw=black!90,
        fill=violet!5,minimum width=0.6cm, minimum height=0.6cm},
    skip/.style={line width=0.2mm, ->},
]
    \def\yshift{0.3em}
    \def\base{8cm}
    \def\dec{0.55cm}

    \node (base) at (0, 0) {};
    \node (pad) at (0, -1cm) {};
    \node (conv) [conv, minimum width=\base - \dec, anchor=south] at
      (base.north) {$\ReLU(\Conv(C_{in}, C_{out}, K=8, S=4))$};
    \node (rewrite) [rewrite, minimum width=\base - 2 * \dec, anchor=south] at
      ($(conv.north) + (0,\yshift)$) {$\GLU(\Conv(2 C_{out}, 2 C_{out}, K=1, S=1))$};

    \def\yshift{0.6em}
    \node (skip) [inout, anchor=east] at ($(rewrite.west) - (\yshift, 0)$) {$\mathrm{Decoder}_i$};
    \draw[->] (rewrite.west) -- (skip);

    \node (prev) [inout, anchor=north] at ($(conv.south) - (0, \yshift)$) {$\mathrm{Encoder}_{i-1}$ or input};
    \draw[->]  (prev.north) -- (conv.south);

    \node (next) [inout, anchor=south] at ($(rewrite.north) + (0, \yshift)$) {$\mathrm{Encoder}_{i+1}$ or LSTM};
    \draw[->]  (rewrite.north) -- (next.south);
\end{tikzpicture}

    \caption{Detailed view of the layers $\mathrm{Decoder}_i$ on the top and
      $\mathrm{Encoder}_i$ on the bottom. Arrows represent connections
    to other parts of the model.}
  \end{subfigure}
  \caption{Demucs complete architecture on the left,
    with detailed representation of the encoder and decoder layers
  on the right.
  Key novelties compared to the previous Wave-U-Net are the GLU activation
  in the encoder and decoder, the bidirectional LSTM in-between
  and exponentially growing number of channels, allowed by the stride of 4
  in all convolutions.}
  \vspace{-0.15cm}
\end{figure}

Our network architecture
is a blend of ideas from the SING architecture~\cite{sing}
developed for music note synthesis and Wave-U-Net.
We reuse the synthesis with large strides and large number of channels as well
as the combination of a LSTM and convolutional layers from SING,
while retaining the U-Net~\cite{unet} structure of Wave-U-Net. The model is composed
of a convolutional encoder, an LSTM and a convolutional decoder, with
the encoder and decoder linked with skip U-Net connections.
The model takes a stereo mixture $s = \sum_i s_i$ as input
and outputs a stereo estimate $\hat{s}_i$ for each source.
Similarly to other work in generation in both image~\cite{gan_style,progressive_gan} and sound~\cite{sing},
we do not use batch normalization~\cite{batchnorm} as our early experiments
showed that it was detrimental to the model performance.

The encoder is composed of $L := 6$ stacked layers numbered from 1 to $L$.
Layer $i$ is composed of
a convolution with kernel size $K:=8$,  stride $S:=4$, $C_{i-1}$ input channels,
$C_{i}$ output channels and ReLU activation
followed by a 1x1 convolution with GLU activation~\cite{glu}.
As the GLU outputs $C/2$ channels with $C$ channels as input,
we double the number of channels in the 1x1 convolution.
We define $C_0 := 2$ the number of channels in the input mixture and $C_1:=48$
the initial number of channels for our model. For $i \in \{2, \ldots, L\}$
we take $C_i := 2 C_{i-1}$ so that the final number of channels is $C_L = 1536$.
We then use a bidirectional LSTM with 2 layers and a hidden size $C_L$.
The LSTM outputs $2 C_L$ channels per time position. We use a 1x1 convolution
with ReLU activation to take that number down to $C_L$.

The decoder is almost the symmetric of the encoder.
It is composed of $L$ layers numbered in reverse order from $L$ to $1$.
The $i$-th layer starts with a convolution with kernel size $3$ and stride $1$,
input/output channels $C_i$ and a ReLU activation.
We concatenate its result with the output of the $i$-th layer of the encoder
to form a U-Net and take back the number of channels to $C_i$ using a 1x1
convolution with GLU activation. Finally, we use a transposed convolution
with kernel size $K= 8$ and stride $S = 4$, $C_{i-1}$ outputs and ReLU activation.
For the final layer, we instead output $4 C_0$ channels and do not use
any activation function.

\vspace{-0.15cm}
\paragraph{Weights rescaling}

The weights of a convolutional layer in a deep learning model are usually
initialized in a way that account for the number of input channels and receptive field of the convolutions (i.e., fan in), as introduced by He et al.~\cite{kaiming_init}. The initial weights of a convolution will roughly
scale as $\frac{1}{\sqrt{K C_{in}}}$ where $K$ is the kernel size and $C_{in}$ the number of input channels.
For instance, the standard deviation
after initialization of the weights of the first layer of our encoder is
about 0.2, while that of the last layer is 0.01.  Modern optimizers such as Adam~\cite{adam} normalize the gradient
update per coordinate so that, on average, each weight will receive
updates of the same magnitude. Thus, if we want to take a learning rate large
enough to tune the weights of the first layer, it will most likely
be too large for the last layer.

In order to remedy this problem, we use a trick that is equivalent to using
specific learning rates per layer. Let us denote $w$ the weights at initialization used
to compute the convolution $w * x$. We take $\alpha := \mathrm{std}(w)/a$,
where $a$ is a reference scale. We replace $w$ by $w' = w / \sqrt{\alpha}$
and the output of the convolution by $\sqrt{\alpha} w' * x$, so that the
output of the layer is unchanged.
This is similar to the equalized learning rate trick used for image generation with GAN~\cite{gan_style}.
We observed both faster decay of the training loss and convergence to a better
optimum when using the weight rescaling trick, see Section~\ref{sec:ablation}.
Optimal performance was obtained for a reference level $a := 0.1$. We also tried rescaling the weights by $1/\alpha$ rather than $1 / \sqrt{\alpha}$ however this made the training loss diverge.

\vspace{-0.15cm}
\paragraph{Synthesis vs.~filtering}

Let us denote $e_i(s)$ the output of the $i$-th layer of the encoder and
$d_i(s)$ the output of the $i$-th layer of the decoder.
Wave-U-Net takes $d_i(s)$ and upsamples it using linear interpolation.
It then concatenates it with $e_{i-1}(s)$ (with $e_0(s) := s$) and applies a convolution with
a stride of 1 to obtain $d_{i-1}(s)$. Thus, it works by taking a coarse
representation, upsampling it, adding back the fine representation
from $e_{i-1}(s)$
and filtering it out to separate channels.

On the other hand, our model takes $d_i(s)$ and concatenates it with $e_{i}(s)$
and uses a transposed convolution to obtain $d_{i-1}(s)$. A transposed convolution
is different from a linear interpolation upsampling. With a sufficient
number of input channels, it can generate any signal, while a linear upsampling
will generate a signal with higher sampling rate but no high frequencies.
High frequencies are injected using
a U-Net skip connection. Separation is performed by applying various filters
to the obtained signal (aka convolution with a stride of 1).

Thus, Wave-U-Net generates its output by iteratively upsampling,
adding back the high frequency part of the signal from the matching encoder output (or from the input for the last decoder layer) and filtering.
On the other hand, our approach consist in a direct synthesis.
The main benefit of synthesis is that we can use a relatively large stride
in the decoder, thus speeding up the computations and
allowing for a larger number of channels.
We believe this larger number of channels is one of the reasons
for the better performance of our model as shown in Section~\ref{sec:results}.

\section{Unlabeled Data Remixing}
\label{sec:remix}

\begin{figure}
    \centering
    \includegraphics[angle=-90,width=0.9\textwidth]{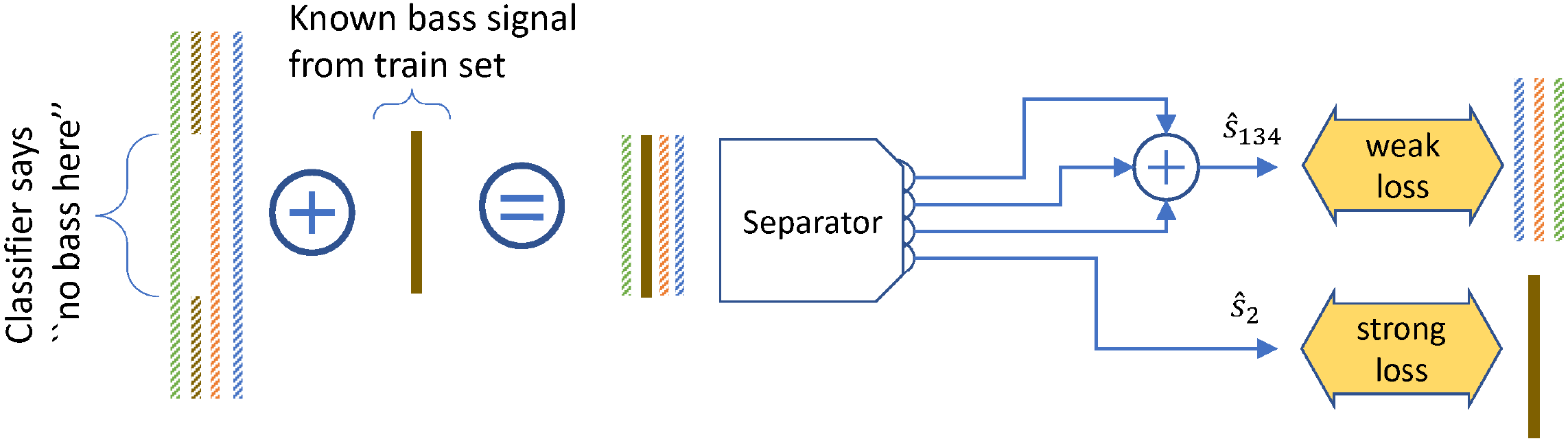}
    \caption{Overall representation of our unlabeled data remixing pipeline.
    When we detect an excerpt of at least 5 seconds with one source silent,
    here the \source{bass}, we recombine it with a single bass sample from the
    training set. We can then provide strong supervision for the silent
    source, and weak supervision for the other 3 as we only know the ground truth
    for their sum.}
    \label{fig:pipeline_remixing}
    \vspace{-0.1cm}
\end{figure}

In order to leverage unlabeled songs, we propose to first train a classifier
to detect the absence or presence of each source on small time frames,
using a supervised train set for which we know the contribution
of each source. When we detect an audio excerpt $m_i$ with
at least 5 seconds of silence for source $i$,
we add it to a new set $\mathcal{D}_i$. We can then mix an
example $m_i \in \mathcal{D}_i$ with a single source $s_i$ taken from the
supervised train set in order to form a new mixture $s' = s_i + m_i$, noting
$s'_j$ the ground truth for this example (potentially unknown to us)
for each source $j$. As the source $i$
is silent in $m_i$ we can provide strong supervision for source $i$ as
we have $s'_i = s_i$ and weak supervision for the other sources as
we have $\sum_{j\neq i} s'_j = m_i$. The whole process pipeline
is represented in Figure~\ref{fig:pipeline_remixing}.

Motivation for this approach comes from our early experiments with the available supervised data
which showed a clear tendency for overfitting when training our separation models.
We first tried using completely
unsupervised regularization, for instance given an unlabeled track $m$, we want $\sum_i \hat{s}_i = m$
where $\hat{s}_i$ is the estimated source $i$.
This proved too weak to improve performance. We then tried to detect passages with a single source present
however this proved too rare of an event in Pop/Rock music: for the standard MusDB dataset presented in
Section~\ref{sec:data}, source \source{other} is alone 2.6\% of the time while the others
are so less than 0.5\% of the time. Accounting for the fact that our model will never reach a recall of 100\%,
this represents too few extractable data to be interesting. On the other hand, a source being silent
happen quite often, 11\% of the time for the \source{drums}, 13\% for the \source{bass} or 32\% for the \source{vocals}.
This time, the \source{other} is the least frequent with 2.6\% and hardest to extract as noted hereafter.

We first formalize our classification problem and then describe the extraction
procedure. The use of the extracted data for training is detailed in Section~\ref{sec:training}.

\subsection{Silent source detector}

\vspace{-0.1cm}
Given a mixture $s = \sum_i s_i$, we define for all sources $i$
the relative volume $V_i(\sob):= 10 \log_{10}\frac{\norm{s_i}^2}{\norm{s}^2}$
and a source being silent as $S_i(\sob) := \mathds{1}_{V_i(\sob) \geq V_{\mathrm{thres}}}$.
For instance, having $V_i = -20$ means that source $i$ is 100 times quieter
than the mixture. Doing informal testing, we observe that
a source with a relative volume between 0 and -10 will be perceived clearly,
between -10 and -20 it will feel like a whisper and almost silent
between -20 and -30. A source with a relative volume under -30 is perceptually
zero.

We can then train a classifier to estimate $P_i := \prob{S_i = 1 | s}$,
the probability that source $i$ is silent given the input mixture $s$.
Given the limited amount of supervised data, we use a Wavelet scattering
transform~\cite{scattering} of order two as input features rather than the
raw waveform. This transformation is computed using the Kymatio package~\cite{kymatio}.
The model is then composed of convolutional layers with
max pooling and batch normalization and a final LSTM that produces an estimate $\hat{P}_i$
for every window of 0.64 seconds with a stride of 64 ms.
We detail the architecture in the Section 2 of the supplementary material.
We have observed that using a downsampled (16kHz) and mono
representation of the mixture further helps prevent overfitting.

The silence threshold is set to -13dB. Although this is far from silent,
this allows for a larger number of positive samples and better training.
We empirically observe that the true relative volumes decreases as
the estimated probability $\hat{P}_i$ increases. Thus, in order to
select only truly silent samples ($V_i \leq -30)$, one only needs to
select a high threshold on $\hat{P}_i$.

\subsection{Extraction procedure}
\label{sec:extraction}

\vspace{-0.1cm}
We assume we have a few labeled data from the same distribution as the unlabeled
data available, in our case we used 100 labeled tracks for 2,000 unlabeled ones,
as explained in Section~\ref{sec:data}. If such data is not available,
it is still possible to annotate part of the unlabeled data, but only with weak labels (source present
or absence) which is easier than obtaining the exact waveform for each source.
We perform extraction by first setting thresholds probabilities $p_i$
for each source. We define $p_i$
as the lowest limit so that for at least 95\% of the samples with $\hat{P}_i(s) \geq p_i$,
we have $V_i(\sob) \leq -20$ on the stem set. We then only keep audio extracts
where $\hat{P}_i \geq p_i$ for at least 5 seconds, which reduces the amount
of data extracted by roughly 50\% but also reduces the 95\% percentile
of the relative volume from -20 to -30. We assemble all the 5 seconds
excerpt where source $i$ is silent into a new dataset $\mathcal{D}_i$.

In our case, we did not manage to obtain more than a few minutes of
audio with source \source{other} silent. Indeed, as noted above,
it is the most frequent source, training examples without it
are rare leading to unreliable prediction.

\section{Experimental results}
\label{sec:results}

\begin{table}
\caption{Comparison of Demucs with the state of the art
in the waveform domain (Wave-U-Net) and in the spectrogram
domain (MMDenseNet, MMDenseNetLSTM) on the MusDB test set.
Extra data is the number of extra songs used, either labeled with
the waveform for each source or unlabeled. We report the median over all tracks
of the median $\SDR$ over each track, as done in the SiSec Mus evaluation campaign~\cite{sisec}.
For easier comparison, the \source{All} column is obtained by concatenating the metrics from all
sources and then taking the median.}
\label{table:comparison}
\begin{center}
  \setlength\extrarowheight{1pt}
\begin{tabular}{l c c c r r r r r}
  \toprule
     && \multicolumn{2}{c}{Extra data} & \multicolumn{5}{c}{Test $\SDR$ in dB}\\
  \textbf{Architecture}& \textbf{Wav?} & \emph{labeled} & \emph{unlabed}  &
  \source{All} & \source{Drums} &  \source{Bass} &\source{Other} & \source{Vocals}\\
  \midrule
  MMDenseNet & \crmark & \crmark & \crmark & 5.34 & 6.40 & 5.14 & 4.13 & 6.57\\

  Wave-U-Net & \chmark & \crmark & \crmark  & 3.17 & 4.16 & 3.17 & 2.24 & 3.05\\
  Demucs & \chmark & \crmark & \crmark     &4.81 & 5.38 & 5.07 & 3.01 & 5.44\\
  \midrule
  Demucs & \chmark & \crmark & 2,000  & 5.09 & 5.79 & 6.23 & 3.45 & 5.51 \\
  Demucs & \chmark & 100 & \crmark  & 5.41 & 5.99 & 5.72 & 3.65 & 6.17 \\
  Demucs & \chmark & 100 & 2,000  & 5.67 & 6.50 & 6.21 & 3.80 & 6.21 \\
  MMDenseLSTM & \crmark & 804 & \crmark & 5.97 & 6.75 & 5.28 & 4.72 & 7.15\\
  MMDenseNet & \crmark & 804 & \crmark  & 5.85 & 6.81 & 5.21 & 4.37 & 6.74\\
  \bottomrule
\end{tabular}
\end{center}
\vspace{-0.2cm}
\end{table}

\begin{table}
\caption{Ablation study for the novel elements in our architecture or training procedure.
Unlike on Table~\ref{table:comparison}, we report the simple $\SDR$ defined in Section~\ref{sec:metrics}
rather than the extended version $\SDR_\ext$.
We also report average values rather than medians as this
make small changes more visible. This explains the $\SDR$ reported here being much smaller than on Table~\ref{table:comparison}.
Both metrics are averaged over the last 3 epochs and computed on the MusDB test set. Reference is trained
with remixed unlabeled data, with stereo channels input resampled at 22kHz, on the train set of MusDB.
}
\label{table:ablation}
\begin{center}
  \setlength\extrarowheight{1pt}
\begin{tabular}{l r r r r}
  \toprule
  & \multicolumn{2}{c}{\textbf{Train set}} & \multicolumn{2}{c}{\textbf{Test set}}\\
  \textbf{Difference}& L1 loss  & $\SDR$ & L1 loss & $\SDR$\\
  \midrule
  Reference & 0.090 & 8.82 &  0.169 & 5.09\\
  no remixed & 0.089 & 8.87  & 0.175 & 4.81\\
  no GLU &   0.099 & 8.00  & 0.174 & 4.68 \\
  no BiLSTM & 0.156 & 8.42   & 0.182 & 4.83\\
  MSE loss  & N/A  & 8.84 & N/A & 5.04\\
  no weight rescaling & 0.094 & 8.39 & 0.171 & 4.68\\

  \bottomrule
\end{tabular}
\end{center}
\vspace{-0.5cm}
\end{table}

We present here the datasets, metrics and baselines used for evaluating our architecture and unlabeled data remixing. We mostly reuse the framework
setup for the SiSec Mus evaluation campaign for music source separation~\cite{sisec} and their MusDB dataset~\cite{musdb}.

\subsection{Evaluation framework}

\paragraph{MusDB and unsupervised datasets}
\label{sec:data}
We use the MusDB~\cite{musdb} dataset, which is composed of 150 songs
with full supervision in stereo and sampled at
44100Hz. For each song, we have the exact waveform of the \source{drums}, \source{bass}, \source{other} and \source{vocals} parts, i.e. each of the sources. The actual song, a.k.a.~the mixture,
is the sum of those four parts. The first 100 songs form the \emph{train set} while the remaining 50 are
kept for the \emph{test set}.

To test out the semi-supervised scheme described in Section~\ref{sec:remix},
we exploit our own set of 2,000 unlabeled tracks,
which represents roughly 4.5 days of audio.
It is composed of 4\% of Heavy Metal, 4\% of Jazz, 37\% of Pop and 55\% of Rock music. Although we do not release
this set, we believe that a similarly composed digital music collection
will allow to replicate our results. We refer to this data as the \emph{unsupervised} or \emph{unlabled} set.

We also collected raw stems for 100 tracks, i.e., individual
instrument recordings used in music production software to make a song.
Those tracks come from the same distribution as our unsupervised dataset but do
not overlap. We manually assigned each instrument to one of the sources using
simple rules on the filenames (for instance ``Lead Guitar.wav'' is assigned to
the \source{other} source) or listening to the stems in case of ambiguity.
We will call this extra supervised data the \emph{stem set}.
As some of the baslises used additional labeled data (807 songs), we also provide metrics for our own architecture trained
using this extra stem set.

We applied our extraction pipeline to the 2,000 unlabeled songs,
and obtained about 1.5 days of audio (with potential overlap due to our extraction procedure) for with the source \source{drums}, \source{bass} or \source{vocals}
silent which form respectively the datasets $\mathcal{D}_0, \mathcal{D}_1, \mathcal{D}_3$.
We could not retrieve a significant amount of audio for the \source{other} source. Indeed, this last source
is the most frequently present (there is almost always a melodic part in a song),
and with the amount of data available, we could not train a model that would
reliably predict the absence of this source. As a consequence, we do not
extract a dataset $\mathcal{D}_2$ for it. We did manage to extract a few hours with only
the \source{other} source, but we have not tried to inject it into our separation model training.
Although we trained our model on mono and 16kHz audio,
we perform the extraction on the original 44kHz stereo data.

\vspace{-0.3cm}
\paragraph{Source separation metrics}
\label{sec:metrics}
Measurements of the performance of source separation models was
developed by Vincent et al.~for blind source separation~\cite{measures}
and reused for supervised source separation in the SiSec Mus evaluation campaign~\cite{sisec}.
Reusing the notations from ~\cite{measures},
let us   take a source $j \in {1, 2, 3, 4}$ and
introduce $P_{s_j}$ (resp $P_\mathbf{s}$)
the orthogonal projection on ${s_j}$ (resp on $\mathrm{Span}(s_1, \ldots, s_4)$).
We then take with $\hat{s}_j$ the estimate of source~$s_j$,
  $s_{\mathrm{target}} := P_{s_j}(\hat{s}_j)$,
  $e_{\mathrm{interf}} := P_{\mathbf{s}}(\hat{s}_j) - P_{s_j}(\hat{s}_j)$ and
  $e_{\mathrm{artif}} := \hat{s}_j - P_{\mathbf{s}}(\hat{s}_j)$.
  The signal to distortion ratio is then defined as
  \begin{equation}
  \SDR := 10 \log_{10}\frac{
      \norm{s_{\mathrm{target}}}^2}{
      \norm{e_{\mathrm{interf}} + e_{\mathrm{artif}}}^2}.
  \end{equation}
Note that this definition is invariant to the scaling of $\hat{s}_j$.
We used the python package
\texttt{museval}\footnote{\url{https://github.com/sigsep/sigsep-mus-eval}}
 which provide a reference implementation for the SiSec Mus 2018 evaluation campaign.
It also allows time invariant filters to be applied
to $\hat{s}_j$ as well as small delays between the estimate and ground truth~\cite{measures}.
As done in the SiSec Mus competition, we report the median over all tracks of the median
of the metric over each track computed using the \texttt{museval} package.
Similarly to previous work~\cite{wavunet,sony_densenet,sony_denselstm}, we focus in this section
on the $\SDR$, but other metrics can be defined ($\SIR$ an $\SAR$) and we present them in the Appendix, Section~\ref{sec:boxplots}.

\vspace{-0.3cm}
\paragraph{Baselines}
We selected the best performing models from the last SiSec Mus evaluation campaign~\cite{sisec} as baselines.
MMDenseNet~\cite{sony_densenet} is a multiscale convolutional network with skip and U-Net connections.
This model was submitted as \textsc{TAK1} when trained without extra labeled data and as \textsc{TAK3}
when trained with 804 extra labeled songs\footnote{Source: \url{https://sisec18.unmix.app/\#/methods/TAK2}}.
MMDenseLSTM~\cite{sony_denselstm} is an extension of MMDenseNet that adds LSTMs at different scales of the encoder and decoder. This model was submitted as \textsc{TAK2} and was trained with the same 804 extra labeled songs. Unlike MMDenseNet, this model was not submitted without supplementary training data.
The only Waveform based method submitted to the evaluation campaign is Wave-U-Net~\cite{wavunet} with the identifier \textsc{STL2}.
Metrics from all baselines were downloaded from the SiSec submission repository\footnote{\url{https://github.com/sigsep/sigsep-mus-2018}}.

\subsection{Training procedure}
\label{sec:training}

We define one epoch over the dataset as a pass over all 5 second extracts with a stride of 0.5 seconds.
We train the classifier described in Section~\ref{sec:remix} on 4 V100 GPUs for 40 epochs
with a batch size of 64 using Adam~\cite{adam} with a learning rate of 5e-4. We use the sum of the binary cross
entropy loss for each source as a training loss.
The Demucs separation model described in Section~\ref{sec:model} is trained for 400 epochs on 16 V100 GPUs,
with a batch size of 128 using Adam with a learning rate of 5e-4 and decaying the learning rate every 160 epochs by
a factor of 5.
We perform the following data augmentation, partially inspired by~\cite{uhlich2017improving}: shuffling sources within one batch,
randomly shifting sources in time (same shift for both channels), randomly swapping channels,
random multiplication by $\pm 1$ per channel. Given the cost of fitting those models, we perform a single run
for each configuration.

We use the L1 distance between the estimated sources $\hat{s}_i$ and the ground truth $s_i$ as we observed
it improved the performance quite a bit, as shown on Table~\ref{table:ablation}. We have tried replacing or adding to this loss
the L1 distance between the power spectrogram of $\hat{s}_i$ and that of $s_i$, as done in~\cite{sing},
however it only degraded the final SDR of the model.
When using the unlabeled data remixing trick describe in Section~\ref{sec:remix}, we perform an extra step with probability
0.25 after each training batch from the main training step. We sample one source $i$ at random out of (0) \source{drums}, (1) \source{bass} or (3) \source{vocals} (remember that we could not extract excerpt for source \source{other}) and obtain
$m_i \in \mathcal{D_i}$ where source $i$ is silent and $s_i$ from the training set where only $i$ is present. We take $s' := m_i + s_i$ and perform a gradient step on the following loss
\begin{equation}
  \norm{\hat{s}'_i -s_i}_1 + \lambda\big\|\sum_{j \neq i} \hat{s}'_j - m_i\big\|_1.
\end{equation}
Given that the extracted examples $m_i$ are noisier than those coming from the train set,
we use a separate instance of Adam for this step with a learning rate 10 times smaller than the main one.
Furthermore, as we only have weak supervision over sources $j\neq i$, the second term is too be understood
as a regularizer rather than a leading term, thus we take $\lambda := 10^{-6}$.

\subsection{Evaluation results}
\label{sec:ablation}

We compare the performance of our approach with the state of the art in Table~\ref{table:comparison}. On the top half,
we show all methods trained without supplementary data. We can see a clear improvement coming from our new architecture
alone compared to Wave-U-Net while MMDenseNet keeps a clear advantage.
We then look at the impact of adding unlabeled remixed data. We obtain a gain of nearly 0.3 of the median SDR. As
a reference, adding 100 labeled tracks to the train set gives a gain of 0.6. Interestingly, even when training
with the extra tracks, our model still benefits from the unlabeled data, gaining an extra 0.2 points of SDR.
  MMDenseLSTM and MMDenseNet still obtain the best performance overall but we can notice that Demucs
trained with unlabeled data achieved state of the art performance for the separation of the \source{bass} source.
It only had access to 100 extra labeled songs, which is far from the 804 extra labeled songs used for MMDenseNet/LSTM
and it would be interesting to see how waveform based models perform with a dataset that large.
Box plots with quantiles can be found in the Appendix, Section~\ref{sec:boxplots}. Audio samples
from different Demucs variant and the baselines are provided online\footnote{\url{https://ai.honu.io/papers/demucs/}}.
We provide an ablation study of the main novelties of this paper on Table~\ref{table:ablation}.
on the train set of MusDB plus our remixed unlabeled data.

\section*{Conclusion}

We presented Demucs, a simple architecture inspired by previous work in source separation from the waveform and
audio synthesis that reduces the gap between spectrogram and waveform based methods from 2.2 points of median SDR
to 0.5 points when trained only on the standard MusDB dataset.
We have also demonstrated how to leverage 2,000 unlabeled mp3s by first training a classifier to detect
excerpt with at least one source silent and then remixing it with an isolated source from the training set.
To our knowledge, this is the first semi-supervised approach to source separation that does not rely on adversarial losses.
Finally, training our model with remixed unlabeled data
as well as 100 extra training examples, we obtain performance almost on par with that of state of the art spectrogram based methods, even better for the \source{bass} source, making waveform based method a legitimate contender for supervised source separation.
\clearpage
\bibliographystyle{plainnat}
\bibliography{references}

\clearpage
\iftoggle{arxiv}{
  \renewcommand{\thesection}{\Alph{section}}

\section*{\Large Appendix}
\setcounter{section}{0}
\renewcommand\theHsection{\Alph{section}}
\section{Architecture of the silent sources detector}

We use as input a scattering transform of order 2, computed using the Kymatio package~\cite{kymatio} with $J:=8$
wavelets per octave. Coefficients of order 1 are indexed by one frequency $f_1$ and of order two by $f_1$ and $f_2$
with $f_2$ the frequency of the second order filter. We organize the coefficient in a tensor of dimension $(C,F,T)$
where $T$ is the number of time windows, $F$ is the number of order 1 frequencies. The first channel is composed
of order 1 coefficients, while the next ones contains the order two coefficient ordered by $f_2$.
Thanks to this reorganization, we can now use 2D convolutions over the output of the scattering transform.
The model is then composed of
\begin{itemize}
    \item batch normalization,
    \item $\ReLU(\Convtwo(C_{in}=7, C_{out}=128, K=5, S=1))$,
    \item $\ReLU(\Convtwo(C_{in}=128, C_{out}=128, K=5, S=1))$,
    \item $\ReLU(\Convtwo(C_{in}=128, C_{out}=256, K=1, S=1))$,
    \item $\mathrm{MaxPool2d}(K=5, S=2)$,
    \item batch normalization,
    \item $\ReLU(\Convtwo(C_{in}=256, C_{out}=256, K=5, S=1))$,
    \item $\ReLU(\Convtwo(C_{in}=256, C_{out}=256, K=5, S=1))$,
    \item $\ReLU(\Convtwo(C_{in}=256, C_{out}=512, K=1, S=1))$,
    \item $\mathrm{MaxPool2d}(K=5, S=2)$,
    \item batch normalization,
    \item frequency dimension is eliminated with a final convolution of kernel size 14 in the frequency axis
        and 1 in the time axis with 512 input channels and 1024 channels as output,
    \item BiLSTM with hidden size of 1024, 2 layers, dropout at 0.18,
    \item $\Conv(C_{in}=2048, C_{out}=1024, K=1, S=1)$,batch normalization, then $\ReLU$,
    \item $\Conv(1024, 4, K=1, S=1)$.
\end{itemize}

\section{Results for all metrics with boxplots}
\label{sec:boxplots}

Reusing the notations from ~\cite{measures},
let us \nico{let us} take a source $j \in {1, 2, 3, 4}$ and
introduce $P_{s_j}$ (resp $P_\mathbf{s}$)
the orthogonal projection on ${s_j}$ (resp on $\mathrm{Span}(s_1, \ldots, s_4)$).
We then take with $\hat{s}_j$ the estimate of source~$s_j$
\begin{align*}
  s_{\mathrm{target}} := P_{s_j}(\hat{s}_j) \qquad
  e_{\mathrm{interf}} := P_{\mathbf{s}}(\hat{s}_j) - P_{s_j}(\hat{s}_j) \qquad
  e_{\mathrm{artif}} := \hat{s}_j - P_{\mathbf{s}}(\hat{s}_j)
\end{align*}
We can now define various signal to noise ratio, expressed in decibels (dB):
the source to distortion ratio
\[
  \SDR := 10 \log_{10}\frac{
      \norm{s_{\mathrm{target}}}^2}{
      \norm{e_{\mathrm{interf}} + e_{\mathrm{artif}}}^2},
\]
the source to interference ratio
\[
  \SIR := 10 \log_{10}\frac{
      \norm{s_{\mathrm{target}}}^2}{
      \norm{e_{\mathrm{interf}}}^2}
\]
and the sources to artifacts ratio
\[
  \SAR := 10 \log_{10}\frac{
    \norm{s_{\mathrm{target}} + e_{\mathrm{interf}}}^2}{
      \norm{e_{\mathrm{artif}}}^2}.
\]

As explained in the main paper, extra invariants are added when using the \texttt{museval} package. We refer
the reader to~\cite{measures} for more details.
We provide hereafter box plots for each metric and each target, generated using
the notebook provided specifically by the organizers of the SiSec Mus evaluation\footnote{\url{https://github.com/sigsep/sigsep-mus-2018-analysis}}. An ``Extra'' suffix means that extra training data has been used and the ``Remixed'' suffix means that our unlabeled data remixing scheme has been used.

\begin{center}
    \includegraphics[width=0.9\textwidth]{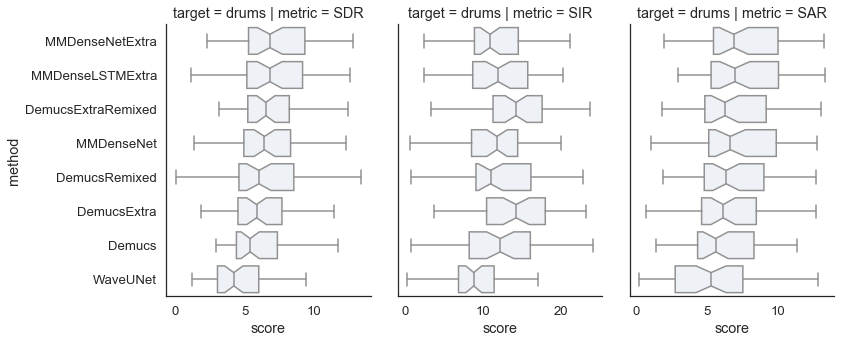}
    \includegraphics[width=0.9\textwidth]{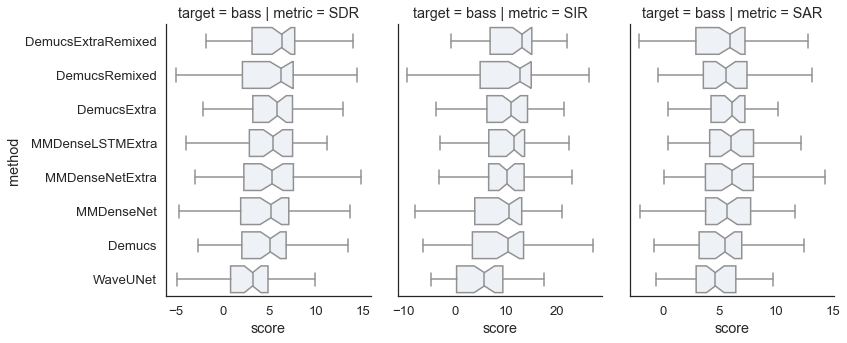}
    \includegraphics[width=0.9\textwidth]{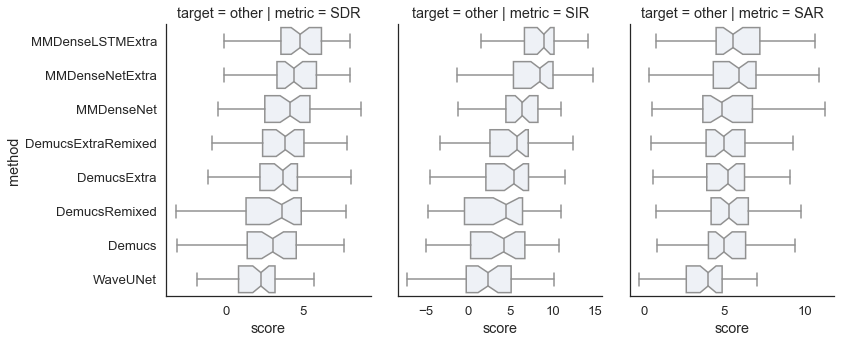}
    \includegraphics[width=0.9\textwidth]{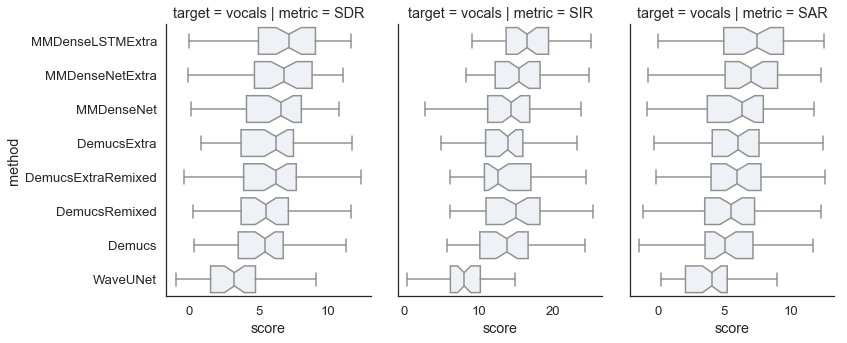}
\end{center}

\end{document}

}{}

\end{document}